\documentclass[12pt, epsfig]{article}
\usepackage{graphicx}
  \usepackage{amsthm,amsfonts}
  \usepackage{amsmath}
\newcommand{\bea}   {\begin{eqnarray}}
\newcommand{\eea}   {\end{eqnarray}}
\begin{document}
\renewcommand{\thefootnote}{\fnsymbol{footnote}}

\thispagestyle{empty}

\title{Effects of Twisted Noncommutativity\\ in Multi-particle Hamiltonians}

\author{Zhanna Kuznetsova\thanks{{\em e-mail: zhanna.kuznetsova@ufabc.edu.br}}
\quad and\quad Francesco
Toppan\thanks{{\em e-mail: toppan@cbpf.br}}
\\
\\
}
\maketitle

\centerline{$^{\ast}$
{\it UFABC, Rua Santa Ad\'elia 166, Bangu,}}{\centerline {\it\quad
cep 09210-170, Santo Andr\'e (SP), Brazil.}
\centerline{$^{\dag}$
{\it CBPF, Rua Dr. Xavier Sigaud 150, Urca,}}{\centerline {\it\quad
cep 22290-180, Rio de Janeiro (RJ), Brazil.}
~\\
\maketitle
\begin{abstract}
~\\
The noncommutativity induced by a Drinfel'd twist produces
Bopp-shift like transformations for deformed operators. In
a single-particle setting the Drinfel'd twist allows to recover the noncommutativity
obtained from various methods which are not based on Hopf algebras. In multi-particle sector, on the other hand, the Drinfel'd twist
implies novel features. In conventional approaches to noncommutativity, deformed primitive operators are postulated to act additively. A Drinfel'd twist implies non-additive
effects which are controlled by the coproduct. We illustrate these features for a class of (abelian twist-deformed) $2D$ Hamiltonians. Suitable choices of the parameters lead to the Hamiltonian of the noncommutative Quantum Hall Effect,
the harmonic oscillator, the quantization of the configuration space.  The non-additive effects in the multi-particle sector, leading to results departing from the existing literature, are pointed out. \end{abstract}
\vfill

\rightline{CBPF-NF-013/12}

\newpage
\section{Introduction}
The aim of this paper is to show that a consistent (non-relativistic) quantization of a Drinfel'd twist \cite{dri} deformed theory implies distinctive features in its multi-particle sector. The resulting, noncommutative, theory is controlled by a Hopf algebra structure \cite{hopf}. Deformed primitive operators acting on multi-particle states are not necessarily additive (the coproduct
unambiguously fixing their non-additive terms). It is rewarding that, in the single-particle sector, the quantization based on the Drinfel'd twist produces the same results obtained from various conventional ways
(not based on a Hopf algebra structure) of dealing with noncommutativity. The departing point with respect to these results lies in the $n$-particle sector (for $n\geq 2$). In most of the conventional approaches, either the problem of constructing multi-particle operators is not addressed or, alternatively, noncommutative primitive operators are postulated to act additively. The Drinfel'd twist induced noncommutativity ($twNC$, for short)
offers a different picture.\par
We illustrate the features of $twNC$ by quantizing a class of (abelian twist-deformed) two-dimensional operators which depend on a set of parameters. By suitably choosing the parameters we can recover, as special cases, the Hamiltonian of the noncommutative Quantum Hall Effect, the $2D$ noncommutative harmonic oscillator, the quantization of the $2D$ configuration space.\par
The quantization framework which respects the Hopf algebra structure and is consistent with the Drinfel'd twist deformation (named ``Unfolded Quantization") was introduced in \cite{ckt} and further discussed in \cite{{cckt},{c2k2t},{top}}. 
In this paper we analyze its model-independent features. The scheme of the paper is as follows.
In Section {\bf 2} we review the basic features of the Unfolded Quantization. In Section {\bf 3} we apply the abelian twist-deformation to a class of operators which include, among others, the harmonic oscillator and the Quantum Hall Effect Hamiltonians. In Section {\bf 4} we compute the effects of the non-additivity in the $2$-particle sector. We comment our results in the Conclusions.

\section{The Unfolded Quantization revisited}

In a non-relativistic system of non-interacting particles certain important operators, including the Hamiltonian, are additive. Indeed, the energy of a system of non-interacting particles is the sum of the energy of its constituents. This feature can be nicely encapsulated within a Hopf algebra framework provided that the additive operators are regarded as primitive operators. This construction, discussed in \cite{ckt}, was named ``Unfolded Quantization". We review here its salient features. \par
At first an abstract dynamical Lie algebra ${\cal G}_D$ is individuated. One of its  generators is associated with the Hamiltonian. In order to allow for the Drinfel'd twist deformations that are discussed in the following, the dynamical Lie algebra ${\cal G}_D$ must contain the Heisenberg-Lie algebra (with generators $x_i, p_i, \hbar$) as a subalgebra. The Universal Enveloping Algebra ${\cal U}({\cal G}_D)$ is naturally endowed with a Hopf algebra structure. In particular a coproduct $\Delta: 
{\cal U}({\cal G}_D)\rightarrow {\cal U}({\cal G}_D)\otimes {\cal U}({\cal G}_D)$ is defined. A physical interpretation of the coproduct is that in the undeformed case it nicely encodes the additivity of the energy. Indeed, let us take any element $g\in {\cal G}_D$. Satisfying the primitivity condition means that $\Delta (g) =g\otimes {\bf 1} +{\bf 1}\otimes g$. For the generator associated to the undeformed Hamiltonian, this condition gets translated into the additivity of the two-particle energy levels, that is $E_{1+2}=E_1+E_2$  (for other physical constructions based on the coproduct see \cite{ls}).\par
The choice of the dynamical Lie algebra ${\cal G}_D$ depends on the properties of the dynamical system under consideration. In this paper we
are focusing on different types of non-relativistic two-dimensional systems which can be derived by a dynamical Lie algebra ${\cal G}_D$ whose
generators can be identified as follows
\bea\label{unfolded0}
{\cal G}_D&=&\{ \hbar, x_1, x_2, p_1,p_2, X_{11}, X_{22}, X_S, P_{11}, P_{22}, P_S, M_{11}, M_{22}, M_{12}, M_{21}\}.
\eea
 The structure constants of the ${\cal G}_D$ Lie algebra can be derived from the structure constants of its  Heisenberg-Lie subalgebra, namely 
\bea
\relax [x_i,p_j]&=&i\hbar \delta_{ij},
\eea
if the following identifications are made
\bea\label{identifications}
&
\begin{array}{llll}
X_{11}=\frac{1}{\hbar}{x_1}^2,&X_{22}=\frac{1}{\hbar}{x_2}^2,&X_{S}=\frac{1}{\hbar}({x_1}x_2+x_2x_1),&\\
P_{11}=\frac{1}{\hbar}{p_1}^2,&P_{22}=\frac{1}{\hbar}{p_2}^2,&P_{S}=\frac{1}{\hbar}({p_1}p_2+p_2p_1),&\\
M_{11}=\frac{1}{\hbar}({x_1}p_1+p_1x_1),&M_{22}=\frac{1}{\hbar}({x_2}p_2+p_2x_2),&&\\
M_{12}=\frac{1}{\hbar}({x_1}p_2+p_2x_1),&M_{21}=\frac{1}{\hbar}({x_2}p_1+p_1x_2).&\\
\end{array}
&
\eea
We deem unnecessary to explicitly present these straighforward computations.\par
Some of the generators entering ${\cal G}_D$ are primitive components of a given Hamiltonian (as discussed in the following), while the remaining ones have to be inserted to guarantee the closure of the ${\cal G}_D$ Lie algebra.
It is important to stress that the generators entering (\ref{unfolded0}) have to be regarded as elements of an abstract Lie algebra.  In particular the Hopf algebra structure of the Universal Enveloping Algebra ${\cal U}({\cal G}_D)$ is only derived from the properties of the abstract Lie algebra ${\cal G}_D$ alone. At the Hopf algebra level the (\ref{identifications}) identifications are inconsistent. \par
The introduction of quantum mechanics further requires the identification of a Hilbert space $V$ which carries a representation of 
${\cal U}({\cal G}_D)$. An operator ${\hat u}=\rho (u)$, acting on $V$, is associated to a given $u\in {\cal U}({\cal G}_D)$. It is recovered from the mapping 
\bea\label{rhomapping}
\rho &:& {\cal U}({\cal G}_D) \rightarrow End(V).
\eea
On $V$, $\hbar$ acts as a multiplication by a constant number. In this paper we set $\hat{\hbar} := \rho(\hbar) ={ 1}$.

\section{A class of operators and their twist-deformations}

We introduce here a class of primitive elements $\Omega\in {\cal U}({\cal G}_D)$, given by the expression
\bea\label{Omega}
\Omega &=& a (P_{11}+P_{22}) + b(X_{11}+X_{22})+c(M_{12}-M_{21})+d x_1+fp_2,
\eea
for $a,b,c,d,f$ arbitrary real parameters. The ${\widehat\Omega}$ operators, obtained via the  (\ref{rhomapping}) transformation,
are Hermitian. By specializing the coefficients entering (\ref{Omega}) we obtain several interesting cases that will be discussed in detail.
In particular we get\par
{\em i}) for $b=1$ and $a=c=d=f=0$, the element $\Omega$ coincides with the ``squared radius"  $R^{2}= X_{11}+X_{22}$;
\par

{\em ii}) for $a=\frac{1}{2}$ and $b=\frac{1}{2}\omega^2$ the element $\Omega$ is associated to the Hamiltonian of the harmonic oscillator
$H=\frac{1}{2}(P_{11}+P_{22})+\frac{1}{2}\omega^2(X_{11}+X_{22})$;\par

{\em iii}) for $a=\frac{1}{2m}$, $b=\frac{m{\omega_c}^2}{8}$, $c=\frac{\omega_c}{4}$, $d=eE$, $f=0$,
we obtain the Quantum Hall Effect Hamiltonian in the presence of constant electric ($E$) and magnetic ($B$) fields ($e$ is the electron's charge,
$\omega_c=\frac{eB}{mc}$ is the cyclotron frequence). \par
The Universal Enveloping Algebra ${\cal U}({\cal G}_D)$ admits an abelian Drinfel'd twist deformation induced by 
${\cal F}\in {\cal U}({\cal G}_D)\otimes {\cal U}({\cal G}_D)$, with
\bea\label{abeliantwist}
{\cal F}^{-1} &=& e^{-i\alpha\epsilon_{ij}p_i\otimes p_j}\equiv {\overline f}^\beta\otimes {\overline f}_\beta
\eea
(in the right hand side, the Sweedler's notation is understood).\par
The deformed generators $\tau^{\cal F}$ induced by the twist (see \cite{{fio},{asc},{ckt}})
\bea
\tau &\mapsto \tau^{\cal F} = {\overline f}^\beta(\tau){\overline f}_\beta,
\eea
generalize the Bopp shift and define, under the ordinary Lie algebra brackets, a non-commutative structure. 
Since, explicitly,
\bea
&\hbar^{\cal F} =\hbar,\quad {p_i}^{\cal F} =p_i,\quad {x_i}^{\cal F} = x_i-\alpha \epsilon_{ij}\hbar p_j,
\eea 
we have that
\bea
\relax [{x_1}^{\cal F},{x_2}^{\cal F}]&=& \Theta=2i\alpha\hbar^2
\eea
(the constant non-commutative parameter $\Theta$ is expressed in terms of the abelian twist parameter $\alpha$).\par
The twist maps $\Omega\in {\cal G}_D$ into ${\Omega}^{\cal F}\in {\cal U}({\cal G}_D)$. We have
\bea\label{omegaF}
\Omega^{\cal F}&=& \Omega + \alpha[ 2b(x_1p_2-p_2x_1)-2c({p_1}^2+{p_2}^2)-d\hbar p_2] +\alpha^2b\hbar ({p_1}^2+{p_2}^2).
\eea
The deformed operator ${\widehat {\Omega}^{\cal F}}\in End(V)$ can be nicely expressed in terms of the creation and annihilation operators
\bea
{a_i}^{(\lambda)} :=\frac{1}{\sqrt{2}}(\lambda {x_i}+i\frac{{ p_i}}{\lambda}) &,&
{{a_i}^{(\lambda)}}^\dagger :=\frac{1}{\sqrt{2}}(\lambda { x_i}-i\frac{{p_i}}{\lambda}),
\eea
for $i=1,2$ and a suitably chosen real parameter $\lambda$ so that
\bea
[a_i^{(\lambda)},{a_j^{(\lambda)}}^\dagger]&=& \delta_{ij}\hbar.
\eea
In all three cases above, namely the deformed squared radius operator, the deformed harmonic oscillator and the deformed Quantum Hall Effect Hamiltonian (in the latter case in the absence of the electric field), the deformed operator ${\widehat \Omega}^{\cal F}$ has the form
\bea\label{defomega}
{\widehat \Omega}^{\cal F}= s(N+1)+tZ\quad\quad (s\geq |t|),
\eea 
in terms of the commuting operators $N,Z$ ($[N,Z]=0$)
\bea
&N={\hat a_1}^\dagger {\hat a_1}+{\hat a_2}^\dagger {\hat a_2},\quad Z= i({\hat a_2}{\hat a_1}^\dagger -{\hat a_1}{\hat a_2}^\dagger),&
\eea
where, for simplicity, the index $\lambda$ in the creation/annihilation operators has been dropped, therefore ${\hat a_i} =\rho(a_i^{(\lambda)})$.
\par
A common pair of eigenvalues $(n,z)$ for the operators $N,Z$ are the integers
$n=0,1,2,\ldots$ and $z=-n+2j$ ($j=0,1,\ldots, n$). It follows that, for $s=|t|$, the vacuum of the operator (\ref{defomega}) is infinitely degenerate.
A unique vacuum solution is obtained when $s>|t|$.\par
The following identifications are obtained:
\par
{\em i}) for the deformed squared radius operator ${\widehat{{R^2}^{\cal F}}}$ we get 
\bea
\lambda =\frac{1}{\sqrt{\alpha}} &,& s=t =2\alpha
\eea
(one should note the singular limit for $\alpha\rightarrow 0$);\par
{\em ii}) for the deformed hamiltonian ${\widehat {H^{\cal F}}}$ of the harmonic oscillator we get
\bea
\lambda ={\sqrt[4]{\frac{\omega^2}{1+\alpha^2\omega^2}}}&,& s=\omega\sqrt{1+\alpha^2\omega^2}, \quad\quad t =\alpha\omega^2;
\eea
{\em iii}) for the deformed hamiltonian ${\widehat {{H_{QHE}}^{\cal F}}}$, in the presence of a constant magnetic field $B$, we get
\bea
\lambda ={\sqrt[4]{\frac{m \omega_c}{2-m\alpha\omega_c}}}&,& s=-t=\frac{1}{2}\omega_c(1-\frac{\alpha \omega_c}{4}).
\eea
In the last two cases the undeformed limit $\alpha\rightarrow 0$ is non-singular.\par
In the third case the turning on of a $\alpha\neq 0$ non-commutativity does not remove the $s=|t|$ degeneracy of the vacuum.\par  
In all three cases a discrete single-particle spectrum of the corresponding deformed operator is recovered from (\ref{defomega}). For the squared radius operator an $\alpha\neq 0$ twist deformation does not just imply a smooth deformation of the undeformed continuum spectrum, but its breaking into a discrete spectrum. \par
In the $s=|t|$ case with an infinitely degenerate vacuum there is a convenient, alternative choice of the creation/annihilation operators, given by
$b,b^\dagger, d, d^\dagger$ satisfying the commutators
\bea
[b,b^\dagger]=[d,d^\dagger]=\hbar^2
\eea
and vanishing otherwise. We recall that, acting on $V$ we have, for the non-vanishing commutators, $[{\hat b},{\hat{b^\dagger}}]=[{\hat d},{\hat{d^\dagger}}]=1$. \par
By setting, for a non-vanishing real $\alpha$,
\bea
b= \frac{1}{2\sqrt{\alpha}}(x_1^{\cal F} +ix_2^{\cal F}), && 
b^\dagger= \frac{1}{2\sqrt{\alpha}}(x_1^{\cal F} -ix_2^{\cal F}),\nonumber\\
d= \frac{1}{2\sqrt{\alpha}}(x_1^{\cal F} -ix_2^{\cal F}+2\alpha\hbar (p_2^{\cal F}+ip_1^{\cal F})), && 
d^\dagger= \frac{1}{2\sqrt{\alpha}}(x_1^{\cal F} +ix_2^{\cal F}+2\alpha\hbar(p_2^{\cal F}-ip_1^{\cal F})),\nonumber\\
\eea
the operator ${\widehat{{R^2}^{\cal F}}}$ can be written as
\bea
{\widehat{{R^2}^{\cal F}}}&=& 2\alpha( 2{\hat {b^\dagger}}{\hat b}+1).
\eea
For the deformed Quantum Hall Effect Hamiltonian ${\widehat {{H_{QHE}}^{\cal F}}}$ (in the presence of the constant electric field $E$) if we define
\bea
{b}=\frac{1}{\sqrt{2}}(a_1-i a_2 +\mu), && b^\dagger = \frac{1}{\sqrt{2}}({a_1}^\dagger +i {a_2}^\dagger +\mu),\nonumber\\
{d}=\frac{1}{\sqrt{2}}(a_1+i a_2 ), && d^\dagger = \frac{1}{\sqrt{2}}({a_1}^\dagger -i {a_2}^\dagger),
\eea
with
\bea
\mu&=& \frac{{\sqrt 2} eE(1-\frac{1}{2}\alpha\omega_c)}{\sqrt{{\omega_c}^3{(1-\frac{1}{4}\alpha\omega_c)}^3}},
\eea
the Hamiltonian can be expressed as
\bea\label{singleQHE}
 {\widehat {{H_{QHE}}^{\cal F}}}&=& W(\hat{b^\dagger}{\hat b}) +K +r({\hat d^\dagger} + {\hat d}),
\eea
where the constants are
\bea\label{WKr}
W&=& \omega_c(1-\frac{1}{4}\alpha\omega_c),\nonumber\\
K&=& \frac{1}{2}\omega_c(1-\frac{1}{4}\alpha\omega_c)- \frac{(eE)^2(1-\frac{1}{2}\alpha\omega_c)^2}{\omega_c^2(1-\frac{1}{4}\alpha\omega_c)^2},\nonumber\\
r&=& \frac{eE}{\sqrt{\omega_c(1-\frac{1}{4}\alpha\omega_c)}}
\eea 
(without loss of generality we have set, for simplicity, $m=\frac{1}{2}$). \par
The spectrum  of ${\widehat {{H_{QHE}}^{\cal F}}}$ is the sum of a discrete contribution proportional to $W$ and of a continuum part proportional to $r$. The latter part vanishes in the absence of the electric field $E$.\par
The present derivation allows to recover the single-particle non-commutative spectrum of the ${\widehat{{R^2}^{\cal F}}}$ operator (see \cite{sghr}), of the harmonic oscillator (see \cite{kk}
and also \cite{oscill}) and of the Quantum Hall Effect Hamiltonian (\cite{dj}, see also \cite{QHE}).

\section{The multi-particle sector of twist-deformed operators}

The multi-particle sector of twist-deformed operators is computed with the help of the coproduct. We discuss at first the $2$-particle state.\par
For any $\tau \in {\cal U}({\cal G}_D)$ the deformed coproduct is defined as
\bea
\Delta^{\cal F}(\tau) &=&{\cal F}\Delta(\tau){\cal F}^{-1} \in {\cal U}({\cal G}_D)\otimes{\cal U}({\cal G}_D).
\eea
Applied on $V\otimes V$, due to the expression  (\ref{abeliantwist}), the twist becomes a unitary operator $F\in End(V\otimes V)$, so that
the $2$-particle operators ${\widehat { \Delta^{\cal F}(\tau)}}$ and ${\widehat { \Delta^{}(\tau)}}$ are unitarily equivalent \cite{{fs},{ckt}}
\bea
{\widehat { \Delta^{\cal F}(\tau)}}&=& F{\widehat { \Delta^{}(\tau)}}F^{-1}\in End({V\otimes V}).
\eea
Therefore, without loss of generality, the deformed $2$-particle operator can be computed from the undeformed coproduct of the deformed element. \par
For $\Omega^{\cal F}$ expressed by (\ref{omegaF})  we have that $ \Delta(\Omega^{\cal F})\in {\cal U}({\cal G}_D)\otimes {\cal U}({\cal G}_D)$ is given by
\bea
\Delta(\Omega^{\cal F})&=& \Delta(\Omega) + 2\alpha b \Delta(x_2p_1-x_1p_2) -2\alpha c \Delta (p_1^2+p_2^2) -\alpha d \Delta (\hbar p_2) +\alpha^2b\Delta (\hbar(p_1^2+p_2^2)).\nonumber\\
&&
\eea
Therefore ${\widehat{\Delta(\Omega^{\cal F})}}\in End(V\otimes V)$
is
\bea\label{nonadditive}
{\widehat{\Delta(\Omega^{\cal F})}}&=&{\widehat {\Omega^{\cal F}}}\otimes {\bf 1}+{\bf 1}\otimes {\widehat {\Omega^{\cal F}}}
+ {\widehat \Omega_r}\otimes {\bf 1}+{\bf 1}\otimes {\widehat\Omega_r} + {\widehat \Omega_{mix}},
\eea
where
\bea
{\widehat \Omega_r}&=& -\alpha d{\hat p_2}+\alpha^2b({\hat p_1}^2+{\hat p_2}^2)\in End(V)
\eea
and
\bea
\relax {\widehat \Omega_{mix}}&=& 2\alpha b[{\hat x_2}\otimes {\hat p_1}+{\hat p_1}\otimes{\hat x_2} -{\hat x_1}\otimes {\hat p_2}-{\hat p_2}\otimes {\hat x_1}]-4\alpha(c-\alpha b)[{\hat p_1}\otimes{\hat p_1} +{\hat p_2}\otimes{\hat p_2}]
\in End(V\otimes V).\nonumber\\
&&
\eea
The non-additivity is implied by the presence of the last three terms in the right hand side of (\ref{nonadditive}).\par
In the absence of the ${\widehat \Omega_{mix}}$ term, the structure of the $2$-particle operator $\widehat{\Delta(\Omega^{\cal F})}$
consists of additive contributions from the first and the second particle. It should be noticed, however, that these contributions ${\widehat\Omega}'$ do not coincide with the deformed $1$-particle operator ${\widehat{\Omega^{\cal F}}}$, due to corrections from the ${\widehat {\Omega}}_r$ term
\bea
{\widehat \Omega}'&=&{\widehat \Omega^{\cal F}}+{\widehat \Omega}_r.
\eea
A convenient formula, for $f=0$, is given by the commutator 
\bea
\relax [ {\widehat\Omega_{mix}}, {\widehat \Omega}' \otimes{\bf {1}}+{\bf 1}\otimes {\widehat\Omega}'] &=&8i\alpha b(c-\alpha b)
[{\hat x_1}\otimes{\hat p_1}+{\hat x_2}\otimes {\hat p_2}+{\hat p_1}\otimes {\hat x}_1+{\hat p_2}\otimes {\hat x}_2]+\nonumber\\
&&
4i\alpha d(c-2\alpha b) [ {\hat p_1}\otimes {\bf 1}+{\bf 1}\otimes {\hat p_1}] - 2i\alpha b d [{\hat x}_2\otimes {\bf 1}+{\bf 1}\otimes {\hat x}_2].
\nonumber\\
&&
\eea
A common set of eigenvalues for ${\widehat \Omega_{mix}}$ and $ {\widehat \Omega}' \otimes{\bf {1}}+{\bf 1}\otimes {\widehat\Omega}'$ is only obtained for $d=0$ and at the special value of $\alpha$ given by $\alpha= \frac{c}{b}$.
\subsection{Center of mass coordinates and $2$-particle eigenvalues of the harmonic oscillator}
Let us discuss now for illustrative purposes the $2$-particle harmonic oscillator Hamiltonian in three cases:
\par
{\em a}) the undeformed $2$-particle Hamiltonian,
\par
{\em b}) the ``additive" noncommutative $2$-particle Hamiltonian constructed as a sum of two $1$-particle noncommutative Hamiltonians and, finally,\par
{\em c}) the twist-deformed $2$-particle Hamiltonian. \par
Let $x_i^{(1)}$, ($x_i^{(2)}$) be the coordinates of the first (second) particle (similarly $p_i^{(1)}$ and $p_i^{(2)}$ are their momenta). The center of mass coordinates and momenta are defined as
\bea
X_i =\frac{1}{2}(x_i^{(1)}+x_i^{(2)}), &&  P_i = p_i^{(1)}+p_i^{(2)},
\eea
while the relative coordinates and momenta are
\bea
y_i =\frac{1}{2}(x_i^{(2)}-x_i^{(1)}), &&  q_i = p_i^{(2)}-p_i^{(1)}.
\eea
The three Hamiltonians are expressed as follows.\par
Case {\em a}:
\bea
H &=& \frac{1}{2} (P_1^2+P_2^2)+2\omega^2(X_1^2+X_2^2) + \frac{1}{2}(q_1^2+q_2^2)+2\omega^2(y_1^2+y_2^2).
\eea  
Case {\em b}:
\bea
H &=& \frac{1}{2} (1+\alpha^2\omega^2)(P_1^2+P_2^2)+2\omega^2(X_1^2+X_2^2) -2\alpha\omega^2(X_1P_2-X_2P_1)+\nonumber\\&& \frac{1}{2}(1+\alpha^2\omega^2)(q_1^2+q_2^2)+2\omega^2(y_1^2+y_2^2)- 2\alpha\omega^2(y_1q_2-y_2q_1).
\eea  
Case {\em c}:
\bea
H &=& (\frac{1}{2}+2\alpha^2\omega^2) (P_1^2+P_2^2)+2\omega^2(X_1^2+X_2^2) -4\alpha\omega^2(X_1P_2-X_2P_1)+ \frac{1}{2}(q_1^2+q_2^2)+2\omega^2(y_1^2+y_2^2).\nonumber\\
&&
\eea  
One can see that in the twist-deformed case the deformation appears only in the center-of-mass sector.\par
The eigenvalues are labeled by the four integers $n_1,n_2,j_1,j_2$ (where $n_1,n_2$ are non-negative, while
$-n_1\leq j_1\leq n_1$, $-n_2\leq j_2\leq n_2$). We have, in the three cases, the following eigenvalues of the energy.
\\
Case {\em a}:
\bea
E_{n_1,n_2,j_1,j_2}&=& 2\omega (n_1+n_2)+4\omega.\eea
Case {\em b}:
\bea
E_{n_1,n_2,j_1,j_2}&=& 2\omega{\sqrt {1+\alpha^2\omega^2} }(n_1+n_2+2)+2\alpha\omega^2(j_1+j_2).\eea
Case {\em c}:
\bea
E_{n_1,n_2,j_1,j_2}&=& 2\omega{\sqrt {1+4\alpha^2\omega^2}} (n_1+1)+2\omega (n_2+1)+4\alpha\omega^2j_1.
\eea
\subsection{The twist deformation of the $2$-particle Quantum Hall Effect Hamiltonian}

The twist-deformed $1$-particle Hamiltonian (\ref{singleQHE}) ${\widehat{{H_{QHE}}^{\cal F}}}$ of the Quantum Hall Effect coincides with the noncommutative Hamiltonian (obtained with other methods) by \cite{dj}. In the $2$-particle sector of the Hamiltonian an extra, non-additive term $H_{extra}$ appears. We have
\bea
{\widehat{{H_{QHE}}^{\cal F}}}^{(2)} &=&{\widehat{{H_{QHE}}^{\cal F}}}\otimes {\bf 1}+{\bf 1}\otimes {\widehat{{H_{QHE}}^{\cal F}}}+H_{extra}.
\eea
The contributions from $H_{extra}$ can be computed, with perturbation method, at the first
order in the expansion of the deformation parameter $\alpha$. We have 
\bea
H_{extra} &=& \alpha H_{(1)}+ O(\alpha^2),
\eea
with
\bea
H_{(1)}&=& -d({\hat p_2}\otimes {\bf 1} +{\bf 1}\otimes {\hat p_2}) + 2b ({\hat x_2}\otimes {\hat p_1}+{\hat p_1}\otimes {\hat x_2} -{\hat x_1}\otimes {\hat p_2}-{\hat p_2}\otimes{\hat x_1}) - 4c ( {\hat p_1}\otimes{\hat p_1}+{\hat p_2}\otimes{\hat p_2}).\nonumber\\&&
\eea
The perturbation theory at the first order gives the energy corrections 
\bea
\Delta E_{(1)} &=& <n_1,n_2;\beta_1,\beta_2|H_{(1)}|n_1,n_2;\beta_1,\beta_2>
\eea
for the $2$-particle states $|n_1,n_2;\beta_1,\beta_2>$ that, in the absence of $H_{extra}$, have
energy eigenvalues
\bea
E_{n_1,n_2;\beta_1,\beta_2}&=& W(n_1+n_2)+ 2K +(\beta_1+\beta_2)r\sqrt{2}
\eea
(the constants $W,K,r$ are given in formula (\ref{WKr})).\par
The computation of the first-order energy corrections $\Delta E_{(1)}$ is lengthy but straightforward. We get, explicitly
\bea\label{extra}
\Delta E_{(1)} &=& \frac{(eE)^2}{8\omega_c} +eE(\beta_1+\beta_2)\sqrt{\frac{\omega_c}{8}}.
\eea
This formula shows the effect of the non-additivity of the Drinfel'd twist deformation in the $2$-particle sector. 
\section{Conclusions}}
This work points out non-additive effects in the multi-particle sector of twist-induced noncommutative nonrelativistic quantum mechanical systems. We discussed a general class of 
operators which, for different choices of their parameters, lead, in particular, to the harmonic oscillator Hamiltonian, the Quantum Hall Effect Hamiltonian and the square distance operator. 
The method based on deformed Hopf algebras allows to recover the single-particle non-commutative
spectrum computed in \cite{kk}, \cite{dj} and \cite{sghr}, respectively. Our work addresses the question of the non-commutative contribution to the multi-particle sector. For the above cases the Drinfel'd twist induced noncommutativity gives definite results, that have been here presented. 
In the case of the harmonic oscillator, non-additive properties of multi-particle systems were derived in \cite{cckt}.  In a different approach based on Moyal star-product non-additive effects
have been recently discussed in \cite{fio2}.
\par
It should be mentioned that, for $n$-particle systems with $n\geq 3$, the non-additive effects satisfy the associativity property as a consequence of the coassociativity of the coproduct \cite{top}.\par
The quantization framework for multi-particle systems here presented can be straightforwardly applied to other types of non-commutative deformations (in particular to non-abelian twists) and is left for future investigations.
~
\\ {~}~
\par {\Large{\bf Acknowledgments}}
{}~\par{}~\par
We are grateful to B. Chakraborty, G. Fiore and F. Scholtz for helpful discussions and to NITheP
(where this paper was initiated) for hospitality. The work received support from CNPq.

\end{document}